\documentclass{ws-mpla}
\usepackage{amsmath,amssymb}

\begin{document}

\markboth{Y. S. Duan, L. D. Zhang and Y. X. Liu} {A New
Description of Cosmic Strings in Brane World Scenario}

\catchline{}{}{}{}{}

\title{A NEW DESCRIPTION OF COSMIC STRINGS IN BRANE WORLD SCENARIO}

\author{
 Yi-Shi Duan,
 Li-Da Zhang\footnote{Corresponding author.
                       Email: zhangld04@lzu.cn},
 Yu-Xiao Liu}
\address{Institute of Theoretical Physics, Lanzhou University,\\
   Lanzhou 730000, P. R. China}

\maketitle

\begin{abstract}
In the light of $\phi$-mapping topological current theory, the
structure of cosmic strings are obtained from the Abelian Higgs
model, which is an effective description to the brane world cosmic
string system. In this topological description of the cosmic
string, combining the result of decomposition of U(1) gauge
potential, we analytically reach the familiar conclusions that in
the brane world scenario the magnetic flux of the cosmic string is
quantized and the RR charge of it is screened.

\end{abstract}

\ccode{PACS Nos.: 11.25.Uv, 11.27.+d, 02.40.-k}





\vspace{7mm}

The current WMAP data has placed the inflationary cosmology on
solid
grounds.\cite{Spergel0603449,Page0603450,Hinshaw0603451,Jarosik0603452}
So the brane inflation\cite{DvaliPLB1999450} has become an
indispensable part of the brane world scenario. In a particularly
simple
version,\cite{BurgessJHEP200107,Dvali0105203,AlexanderPRD200265,JonesJHEP200207,Buchan0311207}
brane inflation takes place when a D3-brane and an anti-D3-brane
move slowly towards each other, and ends as the D3-brane pair
annihilates. Suppose a stack of D3-branes spans our universe, and
the Standard Model fields are open modes on them. When one of them
is anihilated at the end of the inflation, the D-strings as the
cosmic strings will be produced inside (or very close to) the
D3-branes. Assuming all extra dimensions are compactified, the
stability of such D-strings as D1-vortices inside the D3-branes
was demonstrated.\cite{TyeJHEP20040403} It is also pointed out in
Ref. \refcite{TyeJHEP20040403} that the simplest way to see the
existence and stability of D1-vortices in D3-branes is to realize
the direct connection between the D1-vortex solution and a certain
limit of Abelian Higgs (AH) model. So the study of the AH model is
very important to the description of the cosmic strings in the
brane world scenario.

To construct the AH model, the complex scalar Higgs field $\phi$
and U(1) gauge field $A_{\mu}$ inside the D3-brane are introduced.
The phase of the Higgs field $\phi$ is denoted by the axion
$\varphi$ , which is the dual of the RR 2-form field $C_{\mu\nu}$
inside the D3-brane. The axion $\varphi$ measures the winding
number of the D1-vortex and the field $C_{\mu\nu}$ measures the
net RR charge of the D1-vortex. Then the AH model lagrangian is
\begin{equation}
L=-(D^{\mu}\phi)^{*}D_{\mu}\phi-\frac{1}{4}F^{\mu\nu}F_{\mu\nu}-\frac{\lambda}{4}
(\phi^{*}\phi-v^{2})^{2}~~~~(\mu,\nu=0,1,2,3),\label{ah}
\end{equation}
where $D_{\mu}=\partial_{\mu}+ieA_{\mu}$ and
$F_{\mu\nu}=\partial_{\mu} A_{\nu}- \partial_{\nu}A_{\mu}$.

It is well known that the AH model has vortex solution
(Abrikosov-Nielsen-Olesen vortex
\cite{AbrikosovSPJ19575,NielsenNPB197361}). Using so-called
$\phi$-mapping topological current theory, this vortex solution
can be expressed in another way.\cite{DuanJMP200041} And this kind
of expression provides a new description of the cosmic string. In
fact, according to the $\phi$-mapping topological current theory,
the vortex (or string) configurations can always be produced from
the zero points of the complex scalar field in the
four-dimensional space-time. For example, the cosmic string
structures originated from the zero points of the complex scalar
quintessence field were discussed in Ref.
\refcite{DuanJHEP20040402}. The advantage of this description of
the vortex configurations is the physical quantities concerning
the topology of the system can be expressed in an analytical way,
and the relationships of these quantities can be proved strictly.
Therefore, for the purpose of studying the topological property of
the cosmic strings, this description is highly significant.

In the present work, using the $\phi$-mapping topological current
theory, we obtain the cosmic string structures from the AH model,
which is an effective description to the brane world cosmic string
system. Furthermore, combining the result of decomposition of U(1)
gauge potential in Ref. \refcite{DuanJP200214}, we analytically
reach the familiar conclusions that in the brane world scenario
the magnetic flux of the cosmic string is quantized and the RR
charge of it is screened.\cite{TyeJHEP20040403}

To be self-contained basically, we will briefly review the
$\phi$-mapping topological current theory in the following.
Consider a D-dimensional smooth manifold with metric tensor
$g_{\mu\nu}$ and local coordinates $x^{\mu}$. Define a map $\Phi$:
\begin{equation}\label{map}
\Phi^{a}=\Phi^{a}(x^{\mu})~~~(a=1,\cdots,d<D),
\end{equation}
and introduce the direction unit field of $\Phi^{a}$:
\begin{equation}\label{na}
N^{a}=\frac{\Phi^{a}}{\|\Phi\|}~~~(\|\Phi\|=\sqrt{\Phi^{a}\Phi^{a}}).
\end{equation}
Then we can construct a topological tensor current:
\begin{equation}\label{jmn0}
j^{\mu\cdots\nu}\sim\frac{1}{\sqrt{detg_{\mu\nu}}}~\epsilon^{\mu\cdots\nu\lambda\cdots\rho}
 \epsilon^{a\cdots b}\partial_{\lambda}N^{a}\cdots\partial_{\rho}N^{b}.
\end{equation}
It is easy to see that $j^{\mu\cdots\nu}$ is completely
antisymmetric and identically conserved.
 Define the Jacobian tensor $D^{\mu\cdots\nu}(\frac{\Phi}{x})=\frac{1}{d!}
 ~\epsilon^{\mu\cdots\nu\lambda\cdots\rho}
 \epsilon^{a\cdots
 b}\partial_{\lambda}\Phi^{a}\cdots\partial_{\rho}\Phi^{b}$, and the $\phi$-space
Green function $G_{d}(\|\Phi\|)=\left\{%
\begin{array}{ll}
    \frac{1}{\|\Phi\|^{d-2}}, & {d>2;} \\
    \ln\|\Phi\|, & {d=2.} \\
\end{array}%
\right.    $
 Then using
 $\partial_{\mu}N^{a}=\frac{1}{\|\Phi\|}\partial_{\mu}\Phi^{a}+\Phi^{a}\partial_{\mu}
 \frac{1}{\|\Phi\|}$ and the Green function relation $\frac{\partial}{\partial\Phi^{a}}
\frac{\partial}{\partial\Phi^{a}}G_{d}(\|\Phi\|)\sim\delta^{d}(\Phi)$
we can find
\begin{equation}\label{jmn01}
j^{\mu\cdots\nu}\sim\frac{1}{\sqrt{detg_{\mu\nu}}}~\delta^{d}(\Phi)
D^{\mu\cdots\nu}(\frac{\Phi}{x}),
\end{equation}
which implies that $j^{\mu\cdots\nu}\neq0$ only at the points
where $\Phi^{a}=0$. Generally, these points would correspond to
the submanifolds where the topological defects are located. If we
denote the $k$-th above-mentioned submanifold by $M_{k}$, and
define a corresponding normal submanifold $N_{k}$ which is spanned
by the parameter $v^{A}$ $(A=1,\cdots,k)$ with the metric tensor
$g_{AB}$, we can get the only intersection point of $M_{k}$ and
$N_{k}$ denoted by $p_{k}$. By virtue of the implicit function
theorem,\cite{Goursat1904} at the regular points of $\Phi^{a}$,
there exists the only solution to the equations $\Phi^{a}=0$, and
we can expand Eq. (\ref{jmn01}) as
\begin{equation}\label{jmn02}
j^{\mu\cdots\nu}\sim\frac{1}{\sqrt{detg_{\mu\nu}}}~\sum_{k}
\frac{W_{k}\sqrt{detg_{AB}}}{\left.D(\frac{\Phi}{v})\right|_{p_{k}}}~\delta^{d}(M_{k})
D^{\mu\cdots\nu}(\frac{\Phi}{x}),
\end{equation}
where the Jacobian $D(\frac{\Phi}{v})=\frac{1}{d!}
 ~\epsilon^{A\cdots B}
 \epsilon^{a\cdots
 b}\partial_{A}\Phi^{a}\cdots\partial_{B}\Phi^{b}$,
 $\delta^{d}(M_{k})$ is the $\delta$-function on the submanifold $M_{k}$,
  and $W_{k}$ denotes the winding number of the
 $k$-th topological defect. Some important physical quantities
 concerning the topology, say magnetic flux, can be calculated from Eq.
 (\ref{jmn02}). According to the implicit function theorem,\cite{Goursat1904}
 the irregular points of $\Phi^{a}$ correspond to the branch
 points of the topological current $j^{\mu\cdots\nu}$, and the
 branch processes of $j^{\mu\cdots\nu}$ occur at these very
 points. These branch processes can describe various evolutions of
 the topological defects, and the total topological charge of the
 system will keep unchange during these evolutions since $j^{\mu\cdots\nu}$
 is a conserved current.

Before entering into the concrete analysis of the AH model
(\ref{ah}), there remains two more issues that are worth noting.
First, our description of the cosmic string is different from the
exact physical picture of it in a subtle way, since the starting
action (\ref{ah}) is different from the action used in Ref.
\refcite{TyeJHEP20040403} where a $\delta$-function source term
for the RR field is introduced. When we discuss the topological
property of the cosmic string, this difference can be ignored.
 Second, as in Ref. \refcite{TyeJHEP20040403}, our study to the cosmic
 string structure originate from the AH model (\ref{ah}), but the
 conclusions we obtain are not limited to it. For example, the
 RR charge of the cosmic string which we will discuss below is
 absent from the original AH model. Actually, our discussions apply
 to the general situation provided by Ref.
 \refcite{TyeJHEP20040403}.

When the spontaneous symmetry breaking takes place in the system
described by the AH model lagrangian (\ref{ah}), the gauge field
$A_{\mu}$ swallows the axion $\varphi$  , which serves as the
Goldstone Boson, and gains mass
\begin{equation}
m_{A}=\sqrt{2}ev. \label{ma}
\end{equation}
Besides, the Higgs boson also gains mass
\begin{equation}
m_{H}=\sqrt{\lambda}v. \label{mh}
\end{equation}
The value of the ratio $\beta=m_{H}^{2}/m_{A}^{2}$ will decide the
force between the vortices,\cite{HindmarshRPP199558} so it is an
important parameter in the study of the branch processes of the
cosmic strings. Define
\begin{eqnarray}
 J_{\mu}&=&-ie[\phi^{*}(D_{\mu}\phi)-(D_{\mu}\phi)^{*}\phi]  \nonumber \\
        &=&2e^{2}A_{\mu}\phi^{*}\phi-ie[\phi^{*}(\partial_{\mu}\phi)
           -(\partial_{\mu}\phi)^{*}\phi].
\label{jmu}
\end{eqnarray}
Then, from (\ref{ah}), we have the equations of motion
\begin{eqnarray}
&&D^{\mu}D_{\mu}\phi-\frac{\lambda}{2}(\phi^{*}\phi-v^{2})\phi=0,
 \label{m1}\\
&&\partial_{\mu}F^{\mu\nu}=J^{\nu} \label{m2}.
\end{eqnarray}
Set $\phi=\phi^{1}+i\phi^{2}$,
$\phi^{*}\phi=\|\phi\|^{2}=\phi^{a}\phi^{a}(a=1,2)$. We can find
\begin{equation}
J_{\mu}=2e^{2}A_{\mu}\|\phi\|^{2}+2e(\phi^{1}\partial_{\mu}\phi^{2}-\phi^{2}\partial_{\mu}\phi^{1}),
\label{jmu1}
\end{equation}
and separate (\ref{m1}) into two real equations
\begin{eqnarray}
\Box\phi^{1}-e\partial^{\mu}A_{\mu}\phi^{2}-2eA^{\mu}\partial_{\mu}\phi^{2}
-e^{2}A^{2}\phi^{1}-\frac{\lambda}{2}(\|\phi\|^{2}-v^{2})\phi^{1}&=&0,
\label{q1}\\
\Box\phi^{2}+e\partial^{\mu}A_{\mu}\phi^{1}+2eA^{\mu}\partial_{\mu}\phi^{1}
-e^{2}A^{2}\phi^{2}-\frac{\lambda}{2}(\|\phi\|^{2}-v^{2})\phi^{2}&=&0,
\label{q2}
\end{eqnarray}
where $\Box=\partial^{\mu}\partial_{\mu}$ and
$A^{2}=A^{\mu}A_{\mu}$. From above two equations, the current
$J^{\mu}$ is found to be conserved:
\begin{equation}
\partial_{\mu}J^{\mu}=0.
\label{con}
\end{equation}
This result is consistent with Eq. (\ref{m2}).

Using the $\phi$-mapping topological current theory, we can
develop the cosmic strings structures from the Higgs field $\phi$.
Introduce the two-dimensional unit vector $n^{a}$ from $\phi^{a}$:
\begin{equation}
n^{a}=\frac{\phi^{a}}{\|\phi\|}. ~~(a=1,2) \label{defna}
\end{equation}
Define the topological tensor current
\begin{equation}
j^{\mu\nu}=-\frac{1}{e} \epsilon^{\mu\nu\lambda\rho}
\epsilon_{ab}\partial_{\lambda}n^{a}\partial_{\rho}n^{b}.\label{jmn}
\end{equation}
Obviously, $j^{\mu\nu}$ is anti-symmetric and identically
conserved. Similar to Ref. \refcite{DuanJHEP20040402}, we can
obtain
\begin{equation}
j^{\mu\nu}=-\frac{2\pi}{e}\delta^{2}(\phi)D^{\mu\nu}(\frac{\phi}{x}),\label{jmn2}
\end{equation}
where $\delta^{2}(\phi)$ is the $\phi$-space $\delta$-function,
which satisfies the Green function relation
$\frac{\partial}{\partial\phi^{a}}
\frac{\partial}{\partial\phi^{a}}ln\|\phi\|=2\pi\delta^{2}(\phi)$,
and $D^{\mu\nu}(\frac{\phi}{x})=
\frac{1}{2}\epsilon^{\mu\nu\lambda\rho} \epsilon_{ab}
\partial_{\lambda}\phi^{a}\partial_{\rho}\phi^{b}$ is the Jacobian
tensor. Assume the normal submanifolds of the world sheets of the
cosmic strings are spanned by the parameter $v^{1}$ and $v^{2}$.
Then from the implicit function theorem,\cite{Goursat1904}
 it follows that under the regular condition
\begin{equation}\label{reg}
D(\frac{\phi}{v})\equiv\frac{1}{2}
 ~\epsilon^{ab}(\frac{\partial\phi^{a}}{\partial v^{1}}\frac{\partial\phi^{b}}{\partial v^{2}}
 -\frac{\partial\phi^{a}}{\partial v^{2}}\frac{\partial\phi^{b}}{\partial v^{1}})\neq0,
\end{equation}
there exists the only solution to $j^{\mu\nu}\neq0$:
\begin{equation}
x^{\mu}=x^{\mu}_{k}(u^{1},u^{2})~~~(k=1,\cdots,l),\label{string}
\end{equation}
which represents $l$ two-dimensional world sheets of the cosmic
string with intrinsic coordinates $u^{1}$ and $u^{2}$. Though we
obtain above cosmic string structures from the Higgs field $\phi$,
 it is not hard to see that the same result can also be obtained
from the more general complex scalar field with its vacuum
manifold $S^{1}$. Furthermore, the present of the gauge field
$A_{\mu}$ will provide the possibility of the finite energy for
the unit length of the cosmic string.

Using  $\delta$-function theory,\cite{Schouten1951} we can expand
(\ref{jmn2}) in a way with more specific topological
meaning\cite{DuanJHEP20040402}
\begin{equation}
j^{\mu\nu}=-\frac{2\pi}{e}D^{\mu\nu}(\frac{\phi}{x})
\sum^{l}_{k=1}
\frac{W_{k}}{\left.D(\frac{\phi}{v})\right|_{p_{k}}}\int_{S_{k}}
\delta^{4}(x^{\mu}-x^{\mu}_{k}(u)) \sqrt{g_{u}}d^{2}u,\label{jmn3}
\end{equation}
where $W_{k}$ is the winding number of the $k$-th cosmic string,
the Jacobian $D(\frac{\phi}{v})$ is defined in Eq. (\ref{reg}),
$p_{k}$ is the intersection point of the $k$-th worldsheet and its
normal submanifold, $\delta^{4}(x^{\mu}-x^{\mu}_{k}(u))$ denotes
the $\delta$-function in the four-dimensional space-time, $S_{k}$
represents the world sheet of the $k$-th cosmic string, and
$\sqrt{g_{u}}d^{2}u$ is the invariant volume element of the world
sheet.  In light of the implicit function
theorem,\cite{Goursat1904} the spatial components of $j^{\mu\nu}$
is
\begin{eqnarray}
 j^{i}=j^{0i}
    &&=-\frac{2\pi}{e}D^{i}(\frac{\phi}{x})
       \sum^{l}_{k=1} \frac{W_{k}}{\left.D(\frac{\phi}{v})\right|_{p_{k}}}
       \int_{S_{k}}\delta^{4}
       (x^{\mu}-x^{\mu}_{k}(u))\sqrt{g_{u}}d^{2}u \nonumber \\
    &&=-\frac{2\pi}{e}D^{i}(\frac{\phi}{x}) \sum^{l}_{k=1}
         \frac{W_{k}}{\left.D(\frac{\phi}{v})\right|_{p_{k}}}
         \int_{L_{k}}\delta^{3}(\vec{x}-\vec{x}(s))ds \cr \cr
    &&=-\frac{2\pi}{e} \sum^{l}_{k=1} W_{k} \frac{dx^{i}}{ds}
        \int_{L_{k}}\delta^{3}(\vec{x}-\vec{x}(s))ds,\label{ji}
\end{eqnarray}
where $D^{i}(\phi/x)=\frac{1}{2}\epsilon^{ijk}\epsilon_{ab}
\partial_{j}\phi^{a}\partial_{k}\phi^{b}~~(i,j,k=1,2,3)$ is the
Jacobian vector, $L_{k}$ represents the $k$-th cosmic string,
$dx^{i}$ is the line element of the $i$-direction, and $ds$ is the
line element along the cosmic string direction.

From (\ref{jmu}), we have
\begin{eqnarray}
 A_{\mu}&=&\frac{i}{2e\phi^{*}\phi}(\phi^{*}\partial_{\nu}\phi-\partial_{\nu}
         \phi^{*}\phi)-\frac{i}{2e\phi^{*}\phi}[\phi^{*}(D_{\nu}\phi)-
         (D_{\nu}\phi)^{*}\phi] \nonumber \\
         &=&-\frac{1}{e}\epsilon_{ab}n^{a}\partial_{\mu}n^{b}
          +\frac{1}{e}\epsilon_{ab}n^{a}D_{\mu}n^{b}. \label{amu}
\end{eqnarray}
According to the result of decomposition of the U(1) gauge
potential,\cite{DuanJP200214} Eq. (\ref{amu}) can be expressed as
\begin{equation}
   A_{\mu}=-\frac{1}{e}\epsilon_{ab}n^{a}\partial_{\mu}n^{b}
           -\frac{1}{e}\partial_{\mu}\Theta,\label{amu2}
\end{equation}
where $\Theta$ is a phase factor. So the corresponding field
tensor is
\begin{equation}
F_{\mu\nu}=-\frac{2}{e}\epsilon_{ab}\partial_{\mu}n^{a}\partial_{\nu}n^{b}.\label{fmn}
\end{equation}
The magnetic field
\begin{equation}
B^{i}=\frac{1}{2}\epsilon^{ijk}F_{jk}
=-\frac{1}{e}\epsilon^{ijk}\epsilon_{ab}\partial_{j}n^{a}\partial_{k}n^{b}=j^{i},\label{bi}
\end{equation}
i.e. the mathematical form of the magnetic field is identical to
that of the spatial component of $j^{\mu\nu}$ defined in
(\ref{ji}). Therefore, using (\ref{ji}) and (\ref{bi}), we can
find the magnetic flux of the $k$th cosmic string is
\begin{equation}
 \Phi_{k}=\int_{\Sigma_{k}}B^{i}d\sigma_{i}
         =-\frac{2\pi}{e}W_{k},\label{flu}
\end{equation}
where $\Sigma_{k}$ is the two-dimensional spatial surface
perpendicular to $L_{k}$. As showed in (\ref{flu}), once more, we
obtain the familiar result that the magnetic flux of the
Abrikosov-Nielsen-Olesen vortex is quantized. The derivation which
has the similar mathematical form can also be found in Ref.
\refcite{DuanJP200214}. It is worth noting that, during the course
of this derivation, there is no need to assume the current
$J^{\mu}$ vanishes and the inner structure of $A_{\mu}$ and
$F_{\mu\nu}$ (or $B^{i}$) are evident.

From Eqs. (\ref{jmu}), (\ref{amu}) and (\ref{amu2}), it is easy to
see that
\begin{equation}
J_{\mu}=-2e\|\phi\|^{2}\partial_{\mu}\Theta.\label{jm3}
\end{equation}
In terms of the RR 2-form field $C_{\mu\nu}$, we
have\cite{TyeJHEP20040403}
\begin{equation}
 \epsilon^{\mu\nu\lambda\rho}\partial_{\nu}C_{\lambda\rho}
 =\frac{1}{m_{A}}J^{\mu}
 =-\frac{2e}{m_{A}}\|\phi\|^{2}\partial_{\mu}\Theta.\label{cmn}
\end{equation}
For the cylindrical symmetry around a certain cosmic string,
$\|\phi\|^{2}$ is only the function of the distance from the
cosmic string. So
\begin{equation}
\oint\epsilon^{\mu\nu\lambda\rho}\partial_{\nu}C_{\lambda\rho}d\theta=0.\label{cmn2}
\end{equation}
Therefore, from Eqs. (\ref{flu}) and (\ref{cmn2}), we can also
draw the conclusion, which has been given in Ref.
\refcite{TyeJHEP20040403}, that the winding number contribution
from $\phi$ cancels the magnetic flux contribution from $A_{\mu}$
so that the net RR charge of the cosmic string is zero. From the
derivation of Eq. (\ref{cmn2}), we can see that the above
conclusion is a topological property of the cosmic string, and
independent of the detail of the system. So, when two cosmic
strings get close enough to each other, and the cylindrical
symmetry of the field distribution around a single cosmic string
is broken, we could expect that the Eq. (\ref{cmn2}) would not
hold for a single cosmic string any more, but the total RR charge
of the cosmic strings would still be zero globally for the
topological reason.

In conclusion, by virtue of $\phi$-mapping topological current
theory, the structure of cosmic strings has been obtained from the
AH model, which provides an effective description to the brane
world cosmic string system. As mentioned before, this description
of the cosmic string is very similar to that of the vortex in the
type-II superconductor, the fact that the cosmic strings is
originated from the zero points of the Higgs field $\phi$ is
manifested obviously. This is different from the exact physical
picture of the cosmic string mentioned in Ref.
\refcite{TyeJHEP20040403}, that the cosmic strings should be
thought as the points where the phase of the Higgs field $\phi$ is
not defined, even with $\|\phi\|=1$ there. In above-mentioned
description, combining the result of decomposition of U(1) gauge
potential, we have reached the familiar conclusions again as
follows: i) the magnetic flux of the cosmic string is quantized,
ii) the RR charge of the cosmic string is screened.

It is easy to see that the approach used in this paper expresses
the inner structures of the physical quantities analytically and
has the highly topological meaning. Using this approach it has
been proved that when the regular condition (\ref{reg}) is fail,
the branch processes of the cosmic strings will occur, and in
these processes the total winding number as the topological charge
of the cosmic strings is conserved.\cite{JiangJMP200041} We can
expect that the generalized $\phi$-mapping topological current
theory\cite{DuanNPB1998514,DuanPTP1999102,LiuPLB2007650} and the
decomposition theory of the gauge potential beyond
U(1)\cite{Duan9910073,LiPLB2000487,DuanMPLA200217,Duan0201018}
will play a more important role in the future study of the cosmic
strings and other topological defects.

\section*{Acknowledgements}
This work was supported by the National Natural Science Foundation
of the People's Republic of China (No. 502-041016) and the
Fundamental Research Fund for Physics and Mathematics of Lanzhou
University (No. Lzu07002).


\begin{thebibliography}{00}

\bibitem{Spergel0603449}
 D. N. Spergel \emph{et. al.},
 {\it Wilkinson Microwave Anisotropy Probe (WMAP) Three Year Results:
 Implications for Cosmology},
 [astro-ph/0603449].

\bibitem{Page0603450}
 L. Page, \emph{et. al.},
 {\it Three Year Wilkinson Microwave Anisotropy Probe (WMAP) Observations:
 Polarization Analysis},
 [astro-ph/0603450].

\bibitem{Hinshaw0603451}
 G. Hinshaw,\emph{et. al.},
 {\it Three-Year Wilkinson Microwave Anisotropy Probe (WMAP) Observations:
 Temperature Analysis},
 [astro-ph/0603451].

\bibitem{Jarosik0603452}
 N. Jarosik, \emph{et. al.},
 {\it Three-Year Wilkinson Microwave Anisotropy Probe (WMAP) Observations:
 Beam Profiles, Data Processing, Radiometer Characterization and
 Systematic Error Limits},
 [astro-ph/0603452].


\bibitem{DvaliPLB1999450}
 G. R. Dvali and S.-H. H. Tye,
 {\em Phys.Lett.} \textbf{B450} (1999) 72, [hep-ph/9812483].

\bibitem{BurgessJHEP200107}
 C. P. Burgess, M. Majumdar, D. Nolte, F. Quevedo, G. Rajesh,
 and R. Zhang,
 {\em JHEP} {\bf 07} (2001) 047, [hep-th/0105204].

\bibitem{Dvali0105203}
 G. R. Dvali, Q. Shafi, and S. Solganik,
 {\it $D$-brane inflation}, [hep-th/0105203].

\bibitem{AlexanderPRD200265}
 S. H. S. Alexander,
 {\em Phys. Rev.} {\bf D65} (2002) 023507, [hep-th/0105032].

\bibitem{JonesJHEP200207}
 N. Jones, H. Stoica, and S.-H. H. Tye,
 {\em JHEP} {\bf 07} (2002) 051, [hep-th/0203163].

\bibitem{Buchan0311207}
 S. Buchan, B. Shlaer, H. Stoica, and S.-H. H. Tye,
 {\it Inter-brane interactions in compact spaces and brane inflation},
 [hep-th/0311207].

\bibitem{TyeJHEP20040403}
 L. Leblond and S.-H. H. Tye,
 {\em JHEP } {\bf 0403} (2004) 055, [hep-th/0402072].

\bibitem{AbrikosovSPJ19575}
 A. Abrikosov,
 {\em Sov. Phys. JETP} {\bf 5} (1957) 1174.

\bibitem{NielsenNPB197361}
 H. B. Nielsen and P. Olesen,
 {\em Nucl. Phys.} {\bf B61} (1973) 45.

\bibitem{DuanJMP200041}
 Y. S. Duan, L. B. FU, and G. Jia,
 {\em J. Math. Phys} {\bf 41} (2000) 4379, [hep-th/9904123].

\bibitem{DuanJHEP20040402}
 Y. S. Duan and X. Liu,
 {\em JHEP} {\bf 0402} (2004) 028, [hep-th/0304146].

\bibitem{HindmarshRPP199558}
 M. B. Hindmarsh and T. W. B. Kibble,
 {\em Rept. Prog. Phys} {\bf 58} (1995) 477, [hep-th/9904123].

\bibitem{Goursat1904}
 \'{E}. Goursat,
 {\it A Course in Mathematical Analysis},
 vol. I, translated by E. R. Hedrich, (Dover, New York, 1904).

\bibitem{Schouten1951}
 J. A. Schouten,
 {\it TensorAnalysis for  Physicist},
 (Clarendon, Oxfort, 1951).

\bibitem{DuanJP200214}
 Y. S. Duan, X. Liu, and P. M. Zhang,
 {\em J. Phys.: Condens. Matter} {\bf 14} (2002) 7941.

\bibitem{JiangJMP200041}
 Y. Jiang and Y. S. Duan,
 {\em J. Math. Phys.} {\bf 41} (2000) 2616,
 [hep-th/9910104].

\bibitem{DuanNPB1998514}
 Y. S. Duan, S. Li, and G. H. Yang,
 {\em Nucl. Phys.} {\bf B514} (1998) 705.

\bibitem{DuanPTP1999102}
 Y. S. Duan, T. Xu, and G. H.  Yang,
 {\em Prog. Theor. Phys.} {\bf 102} (1999) 467.

\bibitem{LiuPLB2007650}
 Y. X. Liu, L. Zhao, Z. B. Cao, and Y. S. Duan,
 {\em Phys. Lett.} {\bf B650} (2007) 286.

\bibitem{Duan9910073}
 Y. S. Duan and L. B. Fu,
 {\em J. Math. Phys.} {\bf 39} (1998) 4344, [hep-th/9910073].

\bibitem{LiPLB2000487}
 S. Li, Y. Zhang, and Z. Y. Zhu,
 {\em Phys. Lett.} {\bf B487} (2000) 201.

\bibitem{DuanMPLA200217}
 Y. S. Duan and P. M. Zhang,
 {\em Mod. Phys. Lett.} {\bf A17} (2002) 2283.

\bibitem{Duan0201018}
 Y. S. Duan, X. Liu, and L. B. Fu,
 {\em Commun. Theor. Phys.} {\bf 40} (2003) 447,
 [math-ph/0201018].


\end{thebibliography}
\end{document}